\documentclass{article}

\usepackage[preprint, nonatbib]{neurips_2025}
\usepackage[round,authoryear]{natbib}
\setcitestyle{authoryear}
\usepackage{graphicx}
\usepackage{amsmath}
\usepackage[font=scriptsize,labelfont=bf]{caption}  
\usepackage{textcomp}
\usepackage{hyperref}
\usepackage{url}
\usepackage{booktabs}
\usepackage{amsfonts}
\usepackage{nicefrac}
\usepackage{microtype}
\usepackage{xcolor}

\title{Multimodal Recurrent Ensembles for Predicting Brain Responses to Naturalistic Movies (Algonauts 2025)}

\author{%
  Semih Eren%
     \textsuperscript{1,2}%
     \thanks{Equal contribution.}
     \thanks{Corresponding author:
       \href{mailto:seren@cbs.mpg.de}{semih.eren@mailbox.tu-dresden.de}} 
       \thanks{Code available at \url{https://github.com/erensemih/Algonauts2025_ModalityRNN}.}%
     \hspace{0.5em}
     \hspace{0.25em}
  \quad
  Deniz Kucukahmetler\textsuperscript{1,3}\footnotemark[1]\quad
  Nico Scherf\textsuperscript{1,4}\\[1ex]
  \textsuperscript{1}Max Planck Institute for Human Cognitive and Brain Sciences, Leipzig, Germany\\
  \textsuperscript{2}TU Dresden, Dresden, Germany\\
  \textsuperscript{3}School for Embedded and Composite AI (SECAI),Dresden/Leipzig, Germany\\
  \textsuperscript{4}Center for Scalable Data Analytics \& AI (ScaDS.AI), Dresden/Leipzig, Germany
}

\begin{document}
\maketitle

\begin{abstract}
  Accurately predicting distributed cortical responses to naturalistic stimuli requires models that integrate visual, auditory and semantic information over time. We present a hierarchical multimodal recurrent ensemble that maps pretrained video, audio, and language embeddings to fMRI time series recorded while four subjects watched almost 80 hours of movies provided by the Algonauts 2025 challenge. Modality-specific bidirectional RNNs encode temporal dynamics; their hidden states are fused and passed to a second recurrent layer, and lightweight subject-specific heads output responses for 1000 cortical parcels. Training relies on a composite MSE–correlation loss and a curriculum that gradually shifts emphasis from early sensory to late association regions. Averaging 100 model variants further boosts robustness. The resulting system ranked third on the competition leaderboard, achieving an overall Pearson r = 0.2094 and the highest single-parcel peak score (mean r = 0.63) among all participants, with particularly strong gains for the most challenging subject (Subject 5). The approach establishes a simple, extensible baseline for future multimodal brain-encoding benchmarks.
\end{abstract}

\section{Introduction}

Understanding how complex, naturalistic stimuli drive human brain activity is a core question in cognitive computational neuroscience \citep{naselaris_encoding_2011,yamins_performance-optimized_2014,kriegeskorte_deep_2015, doerig_neuroconnectionist_2023}. Recent advances in deep learning and systems neuroscience have produced models that map rich stimulus features onto distributed cortical responses. The Algonauts Project 2025 aims at supporting progress in this area by providing a large open dataset and a clear community benchmark for predicting fMRI BOLD signals \citep{boyle_courtois_2023,gifford_algonauts_2025} from naturalistic movie stimuli. With a common dataset, standardized metrics and a collaborative challenge format, the project allows researchers to compare ideas head-to-head and drives methods development towards more accurate, biologically grounded accounts of how the brain interprets the real world.

\textbf{Our contribution.} In this report, we present the encoding model that earned third place in the Algonauts 2025 challenge. Our key contributions are (i) a multimodal RNN architecture that fuses visual, auditory, and textual embeddings from large pretrained models and maps each input sequence to a sequence of BOLD responses for every region of interest (ROI); (ii) a curriculum‑weighted loss that gradually refocuses the training objective from early visual and somatomotor regions to higher‑order areas, mirroring the brain’s hierarchical processing dynamics; and (iii) a 100‑model ensemble whose averaged predictions maximise robustness and accuracy. This approach achieved an overall score of $r = 0.2094$ (correlation between predicted and measured brain responses). It further attained an across‑subject mean single‑parcel peak score of $r = 0.63$, surpassing the nearest competitor’s peak of $r = 0.60$ by an absolute margin of 0.03, and it performed best on the most challenging participant (Subject 5).

\begin{figure}[!ht]
  \centering
  \includegraphics[width=1.0\textwidth]{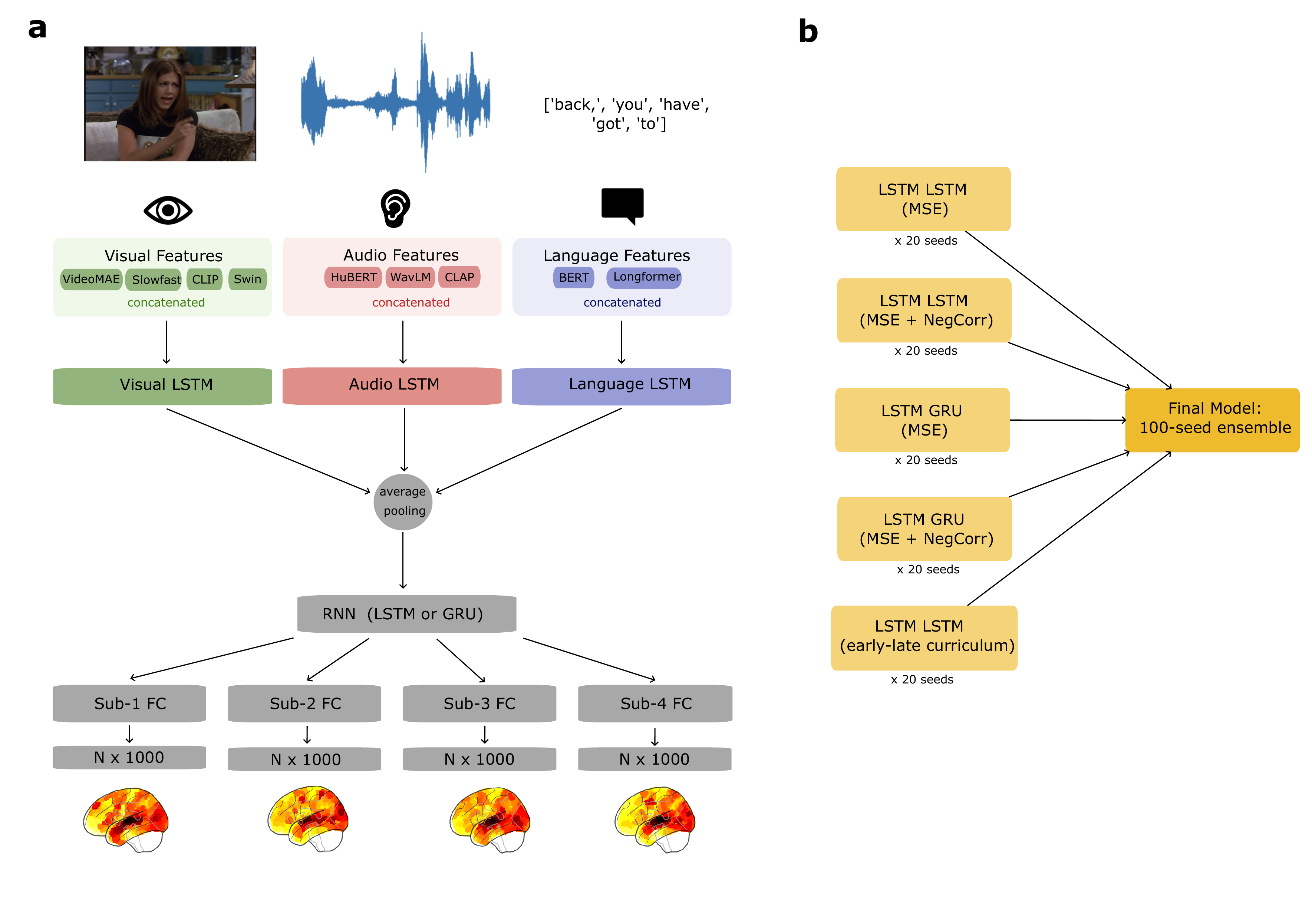}
  \caption{\textbf{a.} Method Illustration. \textbf{b.} Final Model Ensemble Components.}
  \label{fig:impact}
\end{figure}

\section{Related Work}

Recent advances in computational neuroscience have shown that deep learning models provide powerful tools for predicting brain responses to multimodal, naturalistic stimuli \citep{naselaris_encoding_2011,yamins_performance-optimized_2014}. While initial studies primarily encoded fMRI responses to isolated sensory inputs such as static images or speech \citep{kriegeskorte_deep_2015,wen_neural_2018}, recent research highlights the benefits of integrating vision, audio, and language into unified brain-encoding models \citep{gifford_algonauts_2025}. Models combining audiovisual features have notably improved prediction accuracy in higher-order cortical regions \citep{cichy_algonauts_2021}.

Capturing temporal dynamics is essential since natural stimuli induce brain activity evolving over extended timescales. Recurrent neural networks (RNNs), particularly long short-term memory (LSTM) \citep{hochreiter_long_1997} and gated recurrent units (GRUs) \citep{chung_empirical_2014}, have successfully modeled temporal dependencies in brain data. Transformer-based architectures, capable of encoding longer temporal contexts, have also gained attention, though RNN-based methods remain prevalent due to their effectiveness and simplicity \citep{adeli_transformer-based_2023}.

Benchmarking initiatives such as the Algonauts Project significantly accelerated progress through standardized datasets and evaluation methods \citep{gifford_algonauts_2025}. Participants frequently adopt ensemble learning—averaging predictions from multiple diverse models—to enhance accuracy and robustness \citep{nguyen2023algonautsproject2023challenge}. Curriculum learning, progressively shifting training focus from simpler to more complex brain regions, has similarly improved model stability and predictive power \citep{bengio_curriculum_2009}. Our approach integrates these strategies to provide an effective multimodal baseline for future brain-encoding benchmarks.

\section{Algonauts 2025 Challenge and Dataset}

The Algonauts Project is an open challenge platform to bring together researchers in neuroscience and AI to build computational encoding models that accurately predict human brain responses to rich naturalistic stimuli, fostering cross‑disciplinary insights into biological and artificial intelligence. The Algonauts 2025 challenge dataset consists of whole-brain fMRI BOLD responses to naturalistic video stimuli with time-aligned audio and text. It consists of 65 hours of training data (55 hours of Friends seasons 1-6 and Movie10 set which consists of 4 movies). During the model‑building phase, leaderboard scores were based on Friends Season 7, whereas in the model‑selection phase, they were based on a two‑hour‑long out‑of‑distribution movie set. Details can be found at \url{https://algonautsproject.com/challenge.html}.

Neuroimaging data were acquired at a repetition time (TR) of 1.49 s from video clips from movies. Brain responses are summarised into $V=1000$ cortical parcels covering early sensory and higher-order association areas. The evaluation metric is the Pearson correlation between predicted and actual parcel time series averaged across all parcels and subjects.

\section{Method}

\subsection{Model}

Our model has a three‑stage approach that processes multiple input modalities, integrates them into a common representation and predicts subject-specific time‑resolved BOLD signals (Figure 1.a). We extract features from each modality via frozen, pretrained models, feed them into separate LSTMs, average their hidden states and feed those into another RNN that forms a unified latent representation of the multimodal inputs. Finally, this joint embedding is routed through parallel, subject‑specific prediction heads to map latent features to each individual’s brain responses, accounting for personal scaling and idiosyncrasies. In essence, a global model is trained to predict every subject’s brain responses, using an output gating mechanism to direct the shared representation to the appropriate
subject-specific outputs.

\subsubsection{Multimodal Feature Extraction}

To obtain rich representations of each stimulus modality, we leverage state-of-the-art pretrained models. For the visual input, we extract features from each video using four complementary encoders—SlowFast \citep{feichtenhofer_slowfast_2019}, VideoMAE \citep{tong_videomae_2022}, Swin Transformer \citep{liu2021swintransformerhierarchicalvision}, and CLIP \citep{radford2021learningtransferablevisualmodels}—each trained on large-scale video or image datasets. Visual features are computed on 1.49‑second clips and then time‑aligned to the fMRI time points. For the auditory stream, we employ multiple pretrained audio models—HuBERT \citep{hsu_hubert_2021} and WavLM \citep{chen_wavlm_2022} for self‑supervised speech/audio representations, and CLAP \citep{elizalde_clap_2023} for semantic audio embeddings—also extracted from the same 1.49‑second windows and aligned to the fMRI. The text input is represented using two language models applied to dialogue transcripts: a base BERT \citep{devlin_bert_2019} model for local semantic features (by feeding the last $n$ tokens based on the maximum token length of the language model) and a Longformer \citep{beltagy_longformer_2020} for longer‑range context. To ensure continuity of the transcripts of each episode and boost the performance of the first time stamps of the segments (e.g. episode split), we prepend the previous episode's transcripts (if available) when extracting language features.

\subsubsection{Recurrent Modality Encoding and Fusion}

Each modality’s feature sequence $\mathbf{x}_m(t)$ (where $m$ indexes modality) is fed into a dedicated bi‑directional RNN subnetwork (Figure 1.a):
\begin{align}
\mathbf{h}_{m}^{\rightarrow}(t)
  &= \mathrm{RNN}_{m}^{\rightarrow}\bigl(\mathbf{x}_m(t),\,\mathbf{h}_{m}^{\rightarrow}(t-1)\bigr),\nonumber\\
\mathbf{h}_{m}^{\leftarrow}(t)
  &= \mathrm{RNN}_{m}^{\leftarrow}\bigl(\mathbf{x}_m(t),\,\mathbf{h}_{m}^{\leftarrow}(t+1)\bigr),\nonumber\\
\mathbf{h}_m(t)
  &= \bigl[\mathbf{h}_{m}^{\rightarrow}(t)\;;\;\mathbf{h}_{m}^{\leftarrow}(t)\bigr]\in\mathbb{R}^{2H},
\end{align}
where each single‑layer RNN has hidden size $H$ (set to 768).  The concatenated hidden states $\mathbf{h}_m(t)$ form a sequence for each modality.

To combine information across the $M$ modalities, we average them elementwise:
\begin{align}
\bar{\mathbf{h}}(t)
  &= \frac{1}{M}\sum_{m=1}^M \mathbf{h}_m(t),
\end{align}
yielding $\bar{\mathbf{h}}(t)\in\mathbb{R}^{2H}$, an integrated representation at time $t$.  (We found this simple average both effective and regularizing; learned weights or attention did not improve performance.)

The fused sequence $\bar{\mathbf{h}}(t)$ is then fed into a second RNN (denoted as "RNN (LSTM or GRU)" in Figure 1.a), which captures cross‑modality temporal structure:
\begin{align}
\mathbf{z}(t)
  &= \mathrm{RNN}_{\mathrm{post}}\bigl(\bar{\mathbf{h}}(t),\,\mathbf{z}(t-1)\bigr),
\end{align}
producing $\mathbf{z}(t)\in\mathbb{R}^{H}$.

At each time $t$, the final hidden state $\mathbf{z}(t)$ goes into one of four subject‑specific linear heads:
\begin{align}
\mathbf{y}_s(t)
  &= W_s\,\mathbf{z}(t) + b_s,\quad
  W_s\in\mathbb{R}^{1000\times H},\;\mathbf{y}_s(t)\in\mathbb{R}^{1000}.
\end{align}

During training, we pick the head matching the sample’s subject \(s\) and minimize
\begin{equation}
\mathcal{L}
= \frac{1}{T}\sum_{t=1}^T
\Biggl[
\underbrace{\|\mathbf{y}_s(t) - \hat{\mathbf{y}}_s(t)\|_2^2}_{\displaystyle \mathrm{MSE}}
\;-\;
\underbrace{\frac{\sum_{i=1}^N\bigl(y_{s,i}(t)-\bar y_s(t)\bigr)\bigl(\hat y_{s,i}(t)-\overline{\hat y}_s(t)\bigr)}
{\sqrt{\sum_{i=1}^N\bigl(y_{s,i}(t)-\bar y_s(t)\bigr)^2}\;\sqrt{\sum_{i=1}^N\bigl(\hat y_{s,i}(t)-\overline{\hat y}_s(t)\bigr)^2}}}_{\displaystyle r}
\Biggr],
\end{equation}
where \(\mathbf{y}_s(t)\) and \(\hat{\mathbf{y}}_s(t)\) are the model’s predicted and the true fMRI activations over \(N=1000\) parcels, respectively, and
\begin{equation}
\bar y_s(t) = \frac{1}{N}\sum_{i=1}^N y_{s,i}(t),
\qquad
\overline{\hat y}_s(t) = \frac{1}{N}\sum_{i=1}^N \hat y_{s,i}(t).
\end{equation}
At inference, routing through each \(W_s\) yields subject‑specific predictions.  Sharing all recurrent layers but using separate output layers is both parameter‑efficient and sufficiently flexible to capture individual response patterns.

\subsection{Data Cleaning}
To ensure that our subsequent analyses focus only on reliable, informative data, we applied a model‑based filtering step as follows. First, we trained the model on the complete set of movie segments. Next, we used this trained model to generate predictions for each individual segment and computed the performance metric. We then excluded the segments with near-zero correlation scores from the dataset. The result is a pruned collection of segments on which the model can learn and predict with sufficient precision, thereby reducing noise and improving the robustness of downstream results. We removed the Friends season 6 18b, 19a, and 19b segments for subject 1 and Friends season 5 13a and Movie10 bourne01 segment for subject 2.

\subsection{Training}
Using Python 3.10 and PyTorch 2.7, we trained a single sequence‑to‑sequence model in PyTorch with separate subject‑specific heads, using each movie episode paired with its corresponding fMRI time‑series as one sample. To determine the optimal number of epochs, we employed early stopping based on cross‑validation scores. Training was performed with the Adam optimizer at a fixed learning rate of $10^{-3}$ and a batch size of four to ensure balanced subject representation within each batch.

\subsection{Validation}
For validation, we used a group-wise cross-validation scheme, where each fold was defined by an individual movie from Movie10. All of our model improvements were assessed using this framework and throughout the competition our cross-validation score improvements remained closely correlated with our public leaderboard results.

\subsection{"Early-vs-late" curriculum (Loss Weighting)}

One training heuristic we explored was inspired by the hierarchical processing in the brain. Primary sensory areas encode basic stimulus features, whereas higher-order regions accumulate and integrate information over much longer timescales. As a result, low‑level sensory signals tend to be more predictable, while forecasting the responses of higher‑order areas is more challenging. We hypothesised that emphasising early sensory ROIs at the start of training could guide the model to first learn low-level stimulus-response mappings, before focusing on more abstract regions. To implement this, we devised a dynamic loss weighting scheme that gradually shifts focus from early to late-processing ROIs over training epochs. Concretely, we predefined a set of “early-processing” parcels (covering Visual and Somatomotor Network from Schaefer 2018 parcellation \citep{schaefer2018local}) based on the provided ROI labels. During training, we split the loss into two components: $\mathcal{L}_E$ for the early-ROI subset and $\mathcal{L}_L$ for the remaining late-ROI set. We then applied a time-varying weight: at the beginning, the loss weight $w_E$ for $\mathcal{L}_E$ is higher (0.55) and $w_L = 1 - w_E$ is lower (0.45), biasing training towards fitting early ROIs. As epochs progress, we linearly anneal these weights towards a balanced emphasis (e.g. $w_E$ down to 0.5 and $w_L$ up to 0.5). By the end of training, late ROIs receive equal or greater attention. In our experiments, this strategy improved the model’s convergence on challenging ROIs and yielded a small boost in overall correlation performance (especially for mid-level and higher cortical areas) compared to a uniform loss weighting.

\subsection{Ensemble Strategy}

To further improve prediction accuracy, we employed an ensemble of models that achieved similar performances. Besides training models with different random seeds, we increased ensemble diversity by varying the model architecture and training objectives in five distinct ways: (1) using our base RNN architecture with MSE-only loss, (2)  training a variant with a combined MSE and correlation loss, (3) swapping the recurrent units in the post-LSTM to GRU with MSE loss, (4) using the GRU version with combines MSE and correlation loss, (5) using the early-vs-late ROI weighted training curriculum described above (Figure 1.b). 

For each of these five configurations, we trained 20 independent models with different random initialisations (seeds) for 5 epochs. In total, our ensemble comprised 100 models (5 configs × 20 seeds). At prediction time, we averaged the outputs of all ensemble members for each subject and timepoint. This simple averaging yielded a noticeable performance gain (\(\sim \) 2\%) over any single model.

\section{Results and Experiments}

In this section, we summarise the insights that guided our model’s development and ultimately secured third place in the competition with an OOD score of 0.2094. All experiments employ leave‑one‑movie‑out (LOMO) cross‑validation on the Movie10 dataset: for each fold, we train on Friends Seasons 1–6 plus all movies except the held‑out title. We believe LOMO provides a useful diagnostic for evaluating how well our method generalises to unseen data. All of the models are evaluated using the Pearson correlation coefficient as done in the challenge.

\textbf{Model Properties Impact on Results}

We investigated the impact of key architectural choices on our model’s performance by conducting three ablation experiments in comparison to our model with multimodal RNN with subject heads (Figure 2.b). First, Unified modality \& Single Head: we removed both subject-specific output heads and modality-specific encoders—replacing them with a single RNN that processes all modalities and a single prediction head for every subject. Second, Single Head: we retained separate modality encoders but collapsed all subject-specific heads into one shared output head. Third, Unified-Modality: we kept subject-specific heads intact, but replaced all modality-tailored encoders with a single, shared RNN. 

The ablation experiments were successful: our RNN-based modality multi-head model significantly outperformed every ablated variant. (Figure 2.b).

\textbf{Final Models Results}

In Figure~2.c, we present the performance of our five final models—each trained with 20 different random seeds—and their ensemble. Although individual models achieve nearly identical scores, the ensemble consistently outperforms every individual model. Notably, however, in the visual areas the ensemble lags slightly behind our early-late curriculum model, which shows the power of the curriculum approach in the visual areas.

\begin{figure}[!ht]
  \centering
  \includegraphics[width=\textwidth]{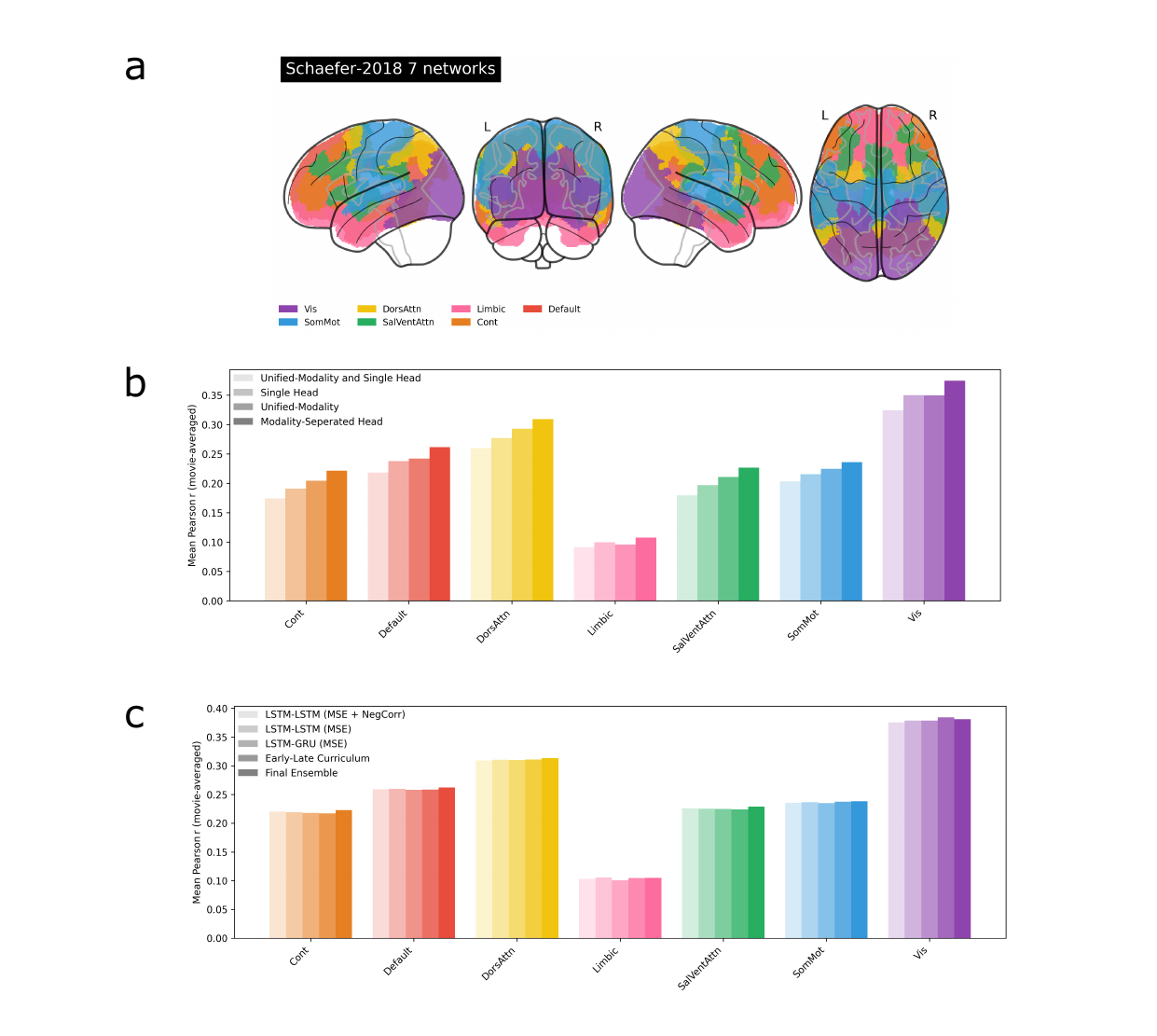}
  \caption{
    (a) Glass‑brain rendering of the Schaefer‑2018 1000‑parcel atlas, highlighting seven functional networks: Visual (Vis), Somatomotor (SomMot), Dorsal Attention (DosAttn), Salience/Ventral Attention (SalVentAttn), Limbic, Frontoparietal (Cont), and Default. Smaller networks are rendered last to ensure full visibility, and the legend maps each color to its corresponding network.
    (b) Ablation study comparing three variants—excluding multi‑head LSTM ("Single Head"), excluding multi‑modal LSTM ("Unified‑Modality"), and excluding both ("Unified‑Modality and Single Head")—against our proposed Modality‑Separated Head model.
    (c) Performance metrics for the final models and their ensemble.
  }
  \label{fig:impact}
\end{figure}

\begin{table}[ht]
  \centering
  \begin{tabular}{lccccc}
    \toprule
    & \textbf{Subject-1} & \textbf{Subject-2} & \textbf{Subject-3} & \textbf{Subject-5} & \textbf{Average} \\
    \midrule
    LSTM-GRU MSE+NegCorr                      & 0.283 & 0.255 & 0.276 & 0.242 & 0.264 \\
    LSTM-LSTM MSE                       & 0.284 & 0.255 & 0.278 & 0.242 & 0.265 \\
    LSTM-GRU MSE                       & 0.284 & 0.254 & 0.278 & 0.241 & 0.264 \\
    LSTM-LSTM MSE+NegCorr                     & 0.283 & 0.254 & 0.277 & 0.242 & 0.264 \\
    Early-late curriculum & 0.285 & 0.254 & 0.279 & 0.242 & 0.265 \\
    \textbf{Ensemble}                                & \textbf{0.289} & \textbf{0.260} & \textbf{0.283} & \textbf{0.247} & \textbf{0.270} \\
    \bottomrule
  \end{tabular}
  \caption{Final Model Ensembling Performances Per Subject}
  \label{tab:results}
\end{table}

\section{Discussion and Limitations}

Our model demonstrates particularly strong performance on the auditory task (as observed from the prediction visualisations on the challenge platform), which we attribute to its multimodal architecture’s ability to leverage a greater number and variety of features heuristically demonstrated in Figure 2.b. Although it does not achieve the highest average score overall, it produces remarkably consistent results across subjects. In an effort to further improve performance and ensemble diversity, we experimented with incorporating a transformer backbone, but it failed to deliver meaningful gains and was therefore abandoned—an omission that may partly explain why we did not obtain the absolute top scores. Notably, while our model achieves the highest peak parcel accuracy, its weakest area lies in the prefrontal cortex (PFC). We hypothesized that this shortcoming stems from insufficient language features, and sought to address it by extending the window of our long-context language feature extractor to include previous episodes, with the goal of capturing more complex, temporally extended brain activity. Although this modification yielded a modest improvement, it did not produce the substantial performance jump we had hoped for. To further address this limitation, we introduced a dynamic ROI‑weighting curriculum inspired by the brain’s hierarchical processing of sensory signals. In this scheme, the model is first encouraged to concentrate its learning on primary sensory ROIs, and only in later stages to reallocate representational weight toward parcels exhibiting more complex, temporally extended dynamics. Although this curriculum produced modest gains in some regions and contributed to the final ensemble, it again fell short of delivering the substantial performance jump we had hoped for in the prefrontal cortex. We further investigated the use of a learning‑rate scheduler. While scheduling afforded substantial gains in single‑model performance, we observed that our seed ensemble without scheduling actually outperformed its scheduled counterpart. Consequently, we opted to use a relatively large fixed learning rate—likely promoting greater ensemble diversity—and achieved superior overall results.

\section{Acknowledgements}

We would like to thank Ariel Iporre Rivas, Pilou Bazin, Katja Seelinger and Kajal Singla for their support and insightful discussions. We also thank the Max Planck Computing and Data Facility (MPCDF) for providing GPU resources. 

D.K. is supported by BMFTR in DAAD project 57616814 (\href{https://secai.org/}{SECAI}).

N.S. is supported by BMBF (Federal Ministry of Education and Research) through
ACONITE (01IS22065) and the Center for Scalable Data Analytics and Artificial Intelligence
(ScaDS.AI.) Leipzig and by the European Union and the Free State of Saxony through BIOWIN.

\bibliographystyle{abbrvnat}

\bibliography{paper}

\begin{thebibliography}{23}
\providecommand{\natexlab}[1]{#1}
\providecommand{\url}[1]{\texttt{#1}}
\expandafter\ifx\csname urlstyle\endcsname\relax
  \providecommand{\doi}[1]{doi: #1}\else
  \providecommand{\doi}{doi: \begingroup \urlstyle{rm}\Url}\fi

\bibitem[Adeli et~al.(2023)Adeli, Minni, and
  Kriegeskorte]{adeli_transformer-based_2023}
H.~Adeli, S.~Minni, and N.~Kriegeskorte.
\newblock Predicting brain activity using transformers.
\newblock \emph{bioRxiv}, 2023.
\newblock \doi{10.1101/2023.08.02.551743}.
\newblock URL
  \url{https://www.biorxiv.org/content/early/2023/08/05/2023.08.02.551743}.

\bibitem[Beltagy et~al.(2020)Beltagy, Peters, and
  Cohan]{beltagy_longformer_2020}
I.~Beltagy, M.~E. Peters, and A.~Cohan.
\newblock Longformer: The long-document transformer, 2020.
\newblock URL \url{https://arxiv.org/abs/2004.05150}.

\bibitem[Bengio et~al.(2009)Bengio, Louradour, Collobert, and
  Weston]{bengio_curriculum_2009}
Y.~Bengio, J.~Louradour, R.~Collobert, and J.~Weston.
\newblock Curriculum learning.
\newblock In \emph{Proceedings of the 26th Annual International Conference on
  Machine Learning}, {ICML} '09, pages 41--48. Association for Computing
  Machinery, 2009.
\newblock ISBN 978-1-60558-516-1.
\newblock \doi{10.1145/1553374.1553380}.
\newblock URL \url{https://doi.org/10.1145/1553374.1553380}.

\bibitem[Boyle et~al.(2023)Boyle, Pinsard, Borghesani, Paugam, {DuPre}, and
  Bellec]{boyle_courtois_2023}
J.~Boyle, B.~Pinsard, V.~Borghesani, F.~Paugam, E.~{DuPre}, and P.~Bellec.
\newblock The courtois {NeuroMod} project: quality assessment of the initial
  data release (2020).
\newblock In \emph{2023 Conference on Cognitive Computational Neuroscience}.
  Cognitive Computational Neuroscience, 2023.
\newblock \doi{10.32470/CCN.2023.1602-0}.
\newblock URL \url{https://2023.ccneuro.org/view_paper2f1e.html?PaperNum=1602}.

\bibitem[Chen et~al.(2022)Chen, Wang, Chen, Wu, Liu, Chen, Li, Kanda, Yoshioka,
  Xiao, Wu, Zhou, Ren, Qian, Qian, Wu, Zeng, Yu, and Wei]{chen_wavlm_2022}
S.~Chen, C.~Wang, Z.~Chen, Y.~Wu, S.~Liu, Z.~Chen, J.~Li, N.~Kanda,
  T.~Yoshioka, X.~Xiao, J.~Wu, L.~Zhou, S.~Ren, Y.~Qian, Y.~Qian, J.~Wu,
  M.~Zeng, X.~Yu, and F.~Wei.
\newblock {WavLM}: Large-scale self-supervised pre-training for full stack
  speech processing.
\newblock \emph{{IEEE} Journal of Selected Topics in Signal Processing},
  16\penalty0 (6):\penalty0 1505--1518, 2022.
\newblock ISSN 1932-4553, 1941-0484.
\newblock \doi{10.1109/JSTSP.2022.3188113}.
\newblock URL \url{http://arxiv.org/abs/2110.13900}.

\bibitem[Chung et~al.(2014)Chung, Gulcehre, Cho, and
  Bengio]{chung_empirical_2014}
J.~Chung, C.~Gulcehre, K.~Cho, and Y.~Bengio.
\newblock Empirical evaluation of gated recurrent neural networks on sequence
  modeling, 2014.
\newblock URL \url{http://arxiv.org/abs/1412.3555}.

\bibitem[Cichy et~al.(2019)Cichy, Roig, Andonian, Dwivedi, Lahner, Lascelles,
  Mohsenzadeh, Ramakrishnan, and Oliva]{cichy_algonauts_2021}
R.~M. Cichy, G.~Roig, A.~Andonian, K.~Dwivedi, B.~Lahner, A.~Lascelles,
  Y.~Mohsenzadeh, K.~Ramakrishnan, and A.~Oliva.
\newblock The algonauts project: A platform for communication between the
  sciences of biological and artificial intelligence, 2019.
\newblock URL \url{http://arxiv.org/abs/1905.05675}.

\bibitem[Devlin et~al.(2019)Devlin, Chang, Lee, and
  Toutanova]{devlin_bert_2019}
J.~Devlin, M.-W. Chang, K.~Lee, and K.~Toutanova.
\newblock {BERT}: Pre-training of deep bidirectional transformers for language
  understanding.
\newblock In J.~Burstein, C.~Doran, and T.~Solorio, editors, \emph{Proceedings
  of the 2019 Conference of the North American Chapter of the Association for
  Computational Linguistics: Human Language Technologies, Volume 1 (Long and
  Short Papers)}, pages 4171--4186. Association for Computational Linguistics,
  2019.
\newblock \doi{10.18653/v1/N19-1423}.
\newblock URL \url{https://aclanthology.org/N19-1423/}.

\bibitem[Doerig et~al.(2023)Doerig, Sommers, Seeliger, Richards, Ismael,
  Lindsay, Kording, Konkle, van Gerven, Kriegeskorte, and
  Kietzmann]{doerig_neuroconnectionist_2023}
A.~Doerig, R.~P. Sommers, K.~Seeliger, B.~Richards, J.~Ismael, G.~W. Lindsay,
  K.~P. Kording, T.~Konkle, M.~A.~J. van Gerven, N.~Kriegeskorte, and T.~C.
  Kietzmann.
\newblock The neuroconnectionist research programme.
\newblock \emph{Nature Reviews Neuroscience}, pages 1--20, 2023.
\newblock \doi{10.1038/s41583-023-00705-w}.

\bibitem[Elizalde et~al.(2022)Elizalde, Deshmukh, Ismail, and
  Wang]{elizalde_clap_2023}
B.~Elizalde, S.~Deshmukh, M.~A. Ismail, and H.~Wang.
\newblock {CLAP}: Learning audio concepts from natural language supervision,
  2022.
\newblock URL \url{http://arxiv.org/abs/2206.04769}.

\bibitem[Feichtenhofer et~al.(2019)Feichtenhofer, Fan, Malik, and
  He]{feichtenhofer_slowfast_2019}
C.~Feichtenhofer, H.~Fan, J.~Malik, and K.~He.
\newblock Slowfast networks for video recognition.
\newblock In \emph{Proceedings of the IEEE International Conference on Computer
  Vision}, pages 6202--6211, 2019.

\bibitem[Gifford et~al.(2025)Gifford, Bersch, St-Laurent, Pinsard, Boyle,
  Bellec, Oliva, Roig, and Cichy]{gifford_algonauts_2025}
A.~T. Gifford, D.~Bersch, M.~St-Laurent, B.~Pinsard, J.~Boyle, L.~Bellec,
  A.~Oliva, G.~Roig, and R.~M. Cichy.
\newblock The algonauts project 2025 challenge: How the human brain makes sense
  of multimodal movies, 2025.
\newblock URL \url{http://arxiv.org/abs/2501.00504}.

\bibitem[Hochreiter and Schmidhuber(1997)]{hochreiter_long_1997}
S.~Hochreiter and J.~Schmidhuber.
\newblock Long short‑term memory.
\newblock \emph{Neural Computation}, 9\penalty0 (8):\penalty0 1735--1780, 1997.

\bibitem[Hsu et~al.(2021)Hsu, Bolte, Tsai, Lakhotia, Salakhutdinov, and
  Mohamed]{hsu_hubert_2021}
W.-N. Hsu, B.~Bolte, Y.-H.~H. Tsai, K.~Lakhotia, R.~Salakhutdinov, and
  A.~Mohamed.
\newblock {HuBERT}: Self-supervised speech representation learning by masked
  prediction of hidden units, 2021.
\newblock URL \url{http://arxiv.org/abs/2106.07447}.

\bibitem[Kriegeskorte(2015)]{kriegeskorte_deep_2015}
N.~Kriegeskorte.
\newblock Deep neural networks: A new framework for modeling biological vision
  and brain information processing.
\newblock \emph{Annual Review of Vision Science}, 1\penalty0 (1):\penalty0
  417--446, 2015.
\newblock ISSN 2374-4642, 2374-4650.
\newblock \doi{10.1146/annurev-vision-082114-035447}.
\newblock URL
  \url{https://www.annualreviews.org/doi/10.1146/annurev-vision-082114-035447}.

\bibitem[Liu et~al.(2021)Liu, Lin, Cao, Hu, Wei, Zhang, Lin, and
  Guo]{liu2021swintransformerhierarchicalvision}
Z.~Liu, Y.~Lin, Y.~Cao, H.~Hu, Y.~Wei, Z.~Zhang, S.~Lin, and B.~Guo.
\newblock Swin transformer: Hierarchical vision transformer using shifted
  windows, 2021.
\newblock URL \url{http://arxiv.org/abs/2103.14030}.

\bibitem[Naselaris et~al.(2011)Naselaris, Kay, Nishimoto, and
  Gallant]{naselaris_encoding_2011}
T.~Naselaris, K.~N. Kay, S.~Nishimoto, and J.~L. Gallant.
\newblock Encoding and decoding in {fMRI}.
\newblock \emph{{NeuroImage}}, 56\penalty0 (2):\penalty0 400--410, 2011.
\newblock ISSN 1095-9572.
\newblock \doi{10.1016/j.neuroimage.2010.07.073}.

\bibitem[Nguyen et~al.(2023)Nguyen, Liu, Li, and
  Luu]{nguyen2023algonautsproject2023challenge}
X.-B. Nguyen, X.~Liu, X.~Li, and K.~Luu.
\newblock The algonauts project 2023 challenge: Uark-ualbany team solution,
  2023.
\newblock URL \url{https://arxiv.org/abs/2308.00262}.

\bibitem[Radford et~al.(2021)Radford, Kim, Hallacy, Ramesh, Goh, Agarwal,
  Sastry, Askell, Mishkin, Clark, Krueger, and
  Sutskever]{radford2021learningtransferablevisualmodels}
A.~Radford, J.~W. Kim, C.~Hallacy, A.~Ramesh, G.~Goh, S.~Agarwal, G.~Sastry,
  A.~Askell, P.~Mishkin, J.~Clark, G.~Krueger, and I.~Sutskever.
\newblock Learning transferable visual models from natural language
  supervision, 2021.
\newblock URL \url{http://arxiv.org/abs/2103.00020}.

\bibitem[Schaefer et~al.(2018)Schaefer, Kong, Gordon, Laumann, Zuo, Holmes,
  Eickhoff, and Yeo]{schaefer2018local}
A.~Schaefer, R.~Kong, E.~M. Gordon, T.~O. Laumann, X.-N. Zuo, A.~J. Holmes,
  S.~B. Eickhoff, and B.~T.~T. Yeo.
\newblock Local-global parcellation of the human cerebral cortex from intrinsic
  functional connectivity {MRI}.
\newblock \emph{Cerebral Cortex (New York, N.Y.: 1991)}, 28\penalty0
  (9):\penalty0 3095--3114, 2018.
\newblock ISSN 1460-2199.
\newblock \doi{10.1093/cercor/bhx179}.

\bibitem[Tong et~al.(2022)Tong, Song, Wang, and Wang]{tong_videomae_2022}
Z.~Tong, Y.~Song, J.~Wang, and L.~Wang.
\newblock {VideoMAE}: Masked autoencoders are data-efficient learners for
  self-supervised video pre-training, 2022.
\newblock URL \url{http://arxiv.org/abs/2203.12602}.

\bibitem[Wen et~al.(2017)Wen, Shi, Zhang, Lu, Cao, and Liu]{wen_neural_2018}
H.~Wen, J.~Shi, Y.~Zhang, K.-H. Lu, J.~Cao, and Z.~Liu.
\newblock Neural encoding and decoding with deep learning for dynamic natural
  vision.
\newblock \emph{Cerebral Cortex}, 28\penalty0 (12):\penalty0 4136--4160, 10
  2017.
\newblock ISSN 1047-3211.
\newblock \doi{10.1093/cercor/bhx268}.
\newblock URL \url{https://doi.org/10.1093/cercor/bhx268}.

\bibitem[Yamins and {DiCarlo}(2016)]{yamins_performance-optimized_2014}
D.~L.~K. Yamins and J.~J. {DiCarlo}.
\newblock Using goal-driven deep learning models to understand sensory cortex.
\newblock \emph{Nature Neuroscience}, 19\penalty0 (3):\penalty0 356--365, 2016.
\newblock ISSN 1546-1726.
\newblock \doi{10.1038/nn.4244}.
\newblock URL \url{https://www.nature.com/articles/nn.4244}.
\newblock Publisher: Nature Publishing Group.

\end{thebibliography}

\end{document}